\documentclass[10pt, conference, compsocconf]{IEEEtran}
\usepackage[T1]{fontenc}
\usepackage{times}
\usepackage[compress]{cite}
\usepackage[dvips]{graphicx}
\usepackage{amsmath}
\usepackage{url}
\usepackage{multirow}
\usepackage{subfigure}
\usepackage{array}
\usepackage{pifont}

\bstctlcite{IEEEexample:BSTcontrol}

\begin{document}

\title{\LARGE \bf Performance Characterization of Multi-threaded Graph Processing Applications on
Many-Integrated-Core Architecture}

\author{Lei Jiang~~~~~Langshi Chen~~~~~Judy Qiu\\
School of Informatics, Computing, and Engineering, Indiana University Bloomington\\
\{jiang60, lc37, xqiu\}@indiana.edu}

\maketitle

\begin{abstract}

In the age of Big Data, parallel graph processing has been a critical technique to analyze and understand connected data. Meanwhile, Moore's Law continues by integrating more cores into a single chip in the deep-nano regime. Many-Integrated-Core (MIC) processors emerge as a promising solution to process large graphs. In this paper, we empirically evaluate various computing platforms including an Intel Xeon E5 CPU, an Nvidia Tesla P40 GPU and a Xeon Phi 7210 MIC processor codenamed Knights Landing (KNL) in the domain of parallel graph processing. We show that the KNL gains encouraging performance and power efficiency when processing graphs, so that it can become an auspicious alternative to traditional CPUs and GPUs. We further characterize the impact of KNL architectural enhancements on the performance of a state-of-the-art graph framework. We have four key observations: \ding{182} Different graph applications require distinctive numbers of threads to reach the peak performance. For the same application, various datasets need even different numbers of threads to achieve the best performance. \ding{183} Not all graph applications actually benefit from high bandwidth MCDRAMs, while some of them favor low latency DDR4 DRAMs. \ding{184} Vector processing units executing AVX512 SIMD instructions on KNLs are underutilized when running the state-of-the-art graph framework. \ding{185} The sub-NUMA cache clustering mode offering the lowest local memory access latency hurts the performance of graph benchmarks that are lack of NUMA awareness. At last, we suggest future works including system auto-tuning tools and graph framework optimizations to fully exploit the potential of KNL for parallel graph processing.
\end{abstract}

\IEEEpeerreviewmaketitle

\section{Introduction}


Big Data explodes exponentially in the form of large-scale graphs. Graph processing has been an important technique to compute, analyze and visualize connected data. Real-world large-scale graphs~\cite{Nai:SC2015,Wang:PPoPP2015} include social networks, transportation networks, citation graphs and cognitive networks, which typically have millions of vertices and millions to billions of edges. Because of the huge graph size, it is natural for both industry and academia to develop parallel graph frameworks to accelerate graph processing on various hardware platforms. A large number of CPU frameworks, e.g., Giraph~\cite{Tian:VLDB2013}, Pregel~\cite{Malewicz:SIGMOD2010} and GraphLab~\cite{Low:VLDB2012}, have emerged for scalable in-memory graph processing. Recent works propose GPU frameworks such as CuSha~\cite{Khorasani:HPDC2014}, Medusa~\cite{Zhong:SIGMOD2014}, LonestarGPU~\cite{Burtscher:IISWC2012} and Gunrock~\cite{Wang:PPoPP2015} to perform high throughput graph processing on GPUs.

In CPU frameworks, graphs are partitioned, distributed and processed among multiple nodes~\cite{Low:VLDB2012,Tian:VLDB2013,Malewicz:SIGMOD2010} by message passing schemes or computed locally on a shared memory node~\cite{Nai:SC2015}. Distributed graph processing is notorious for frequent communications between computations on each vertex or edge~\cite{Nai:SC2015}. In a large-scale cluster, frequent communications are translated to a huge volume of messages across multiple nodes~\cite{Low:VLDB2012,Tian:VLDB2013,Malewicz:SIGMOD2010}, seriously limiting the performance of graph processing. Even a single Mac-Mini SSD-based laptop potentially can outperform a medium-scale cluster for graph processing~\cite{Kyrola:OSDI2012}. On a shared memory multi-core CPU node, communications are interpreted to loads and stores in memory hierarchy~\cite{Nai:SC2015}. The core efficiency of a shared memory node is averagely 100$\times$ higher than that in a cluster~\cite{Suzumura:IISWC2011}. However, compared to GPUs, the graph computing throughput on CPUs is still constrained by the limited number of cores.

GPU frameworks capture graph processing parallelism by mapping and computing vertices or edges on thousands of tiny GPU cores~\cite{Burtscher:IISWC2012,Wang:PPoPP2015}. Although graph applications exhibit irregular memory access patterns, frequent thread synchronizations and data dependent control flows~\cite{Burtscher:IISWC2012}, GPUs still substantially boost the performance of graph processing over CPUs. However, the performance of graph applications on GPUs is sensitive to graph topologies. When traversing graphs with large diameter and small degree, GPUs exhibit poor performance due to the lack of traversal parallelism~\cite{Wang:PPoPP2015}. Moreover, in some commercial applications, simple operations, such as adding or deleting an edge in a graph, may cost a significant portion of the application execution time~\cite{Nai:SC2015}, but it is difficult for GPUs to support such basic operations.

Recently, Intel releases Xeon Phi MIC processors as an alternative to traditional CPUs and GPUs for the high performance computing market. A MIC processor delivers the GPU-comparable computing throughput and CPU-like sequential execution power. Compared to GPUs, a MIC processor can easily support basic graph operations, e.g., inserting/deleting a node/edge, since it is fully compatible with the X86 instruction set. In this paper, we focus on the performance characterization of multi-threaded graph applications on an Intel Xeon Phi (2nd generation) processor - KNL~\cite{Sodani:HOT2015}. Our contributions are summarized as follows:

\begin{itemize}
\item We empirically evaluate a state-of-the-art graph framework on three hardware platforms: an Intel Xeon E5 CPU, an Nvidia Tesla P40 GPU and a KNL MIC processor. We show that the KNL demonstrates encouraging performance and power efficiency when running multi-threaded graph applications. 
\item We further characterize performance details of multi-threaded graph benchmarks on KNL. We measure and analyze the impact of KNL architectural enhancements such as many out-of-order cores, simultaneous multithreading, cache clustering modes, vectorization processing units and 3D stacked MCDRAM on parallel graph processing. We have four key observations: \ding{182} Different graph applications require distinctive numbers of threads to achieve their best performance. For the same benchmark, various datasets ask for different numbers of threads to attain the shortest execution time. \ding{183} Not all graph applications actually benefit from high bandwidth MCDRAMs, while some of them favor low latency DDR4 DRAMs. \ding{184} VPUs executing AVX512 SIMD instructions on KNLs are underutilized when processing graphs. \ding{185} The sub-NUMA cache clustering mode offering the lowest local memory access latency hurts the performance of graph benchmarks that are lack of NUMA awareness.
\item For future work, we suggest possible system auto-tuning tools and framework optimizations to fully exploit the potential of KNL for multi-threaded graph processing applications.
\end{itemize}

\section{Background}
\label{s:back_ground}

\subsection{Graph Processing Frameworks and Benchmark Suite}

The rapid and wide deployment of graph analytics in real-world diversifies graph applications. A lot of applications such as breadth first search incorporate graph traversals, while other benchmarks such as page rank involve intensive computations on vertex properties. Though low-level hardwired implementations~\cite{Beamer:IISWC2015,Ahamd:IISWC2015,Jiang:ICS2016,Chen:CGO2016} have demonstrated high computing efficiency on both CPUs and GPUs, programmers need high-level programmable frameworks to implement a wide variety of complex graph applications to solve real-world problems. Therefore, graph processing heavily depends on specific frameworks composed of data structures, programming models and basic graph primitives to fulfill various functionalities. Previous research efforts propose many graph frameworks on both CPUs~\cite{Nai:SC2015,Low:VLDB2012} and GPUs~\cite{Nai:SC2015,Zhong:SIGMOD2014,Burtscher:IISWC2012,Khorasani:HPDC2014} to hide complicated details of operating on graphs and provide basic graph primitives. Most workloads spend $50\%\sim76\%$ of execution time within graph frameworks~\cite{Nai:SC2015}.

The graph applications we studied in this paper are from a state-of-the-art graph benchmark suite, {\it graphBIG}~\cite{Nai:SC2015}. The suite is implemented with IBM {\it System-G} framework that is a comprehensive graph library used by many industrial solutions~\cite{Tanase:GDMES2014}. The framework adopts the vertex centric programming model, where a vertex is a basic element of a graph. Properties and outgoing edges of a vertex are attached to the same vertex data structure. The data structure of all vertices is stored in an adjacency list, and outgoing edges inside a vertex data structure also form an adjacency list of edges. The {\it major} reason to choose graphBIG is that each benchmark in this suite has a CPU OpenMP version and a GPU CUDA version sharing the same data structure and vertex centric programming model provided by System-G. In this way, we can minimize the impact of differences between CPU and GPU graph processing frameworks, and focus on only differences between various hardware computing platforms. On the contrary, other graph suites~\cite{Beamer:IISWC2015,Ahamd:IISWC2015,Burtscher:IISWC2012} implement graph applications without using a general framework or by using different frameworks on CPUs and GPUs respectively. Particularly, as one of the best-known graph benchmark suites, Graph500~\cite{Murphy:CUG2010} provides only limited number of CPU workloads, since it is designed for only CPU-based system performance ranking.

\vspace{-0.1in}
\subsection{Target Graph Benchmarks}
\vspace{-0.05in}

Graph applications exhibit diversified program behaviors. Some workloads, e.g., breadth first search, are traversal-based. Only a subset of vertices is typically active at a given point during the execution of this type of applications. They often introduce many memory accesses but limited arithmetic operations. Their irregular memory behavior results in extremely poor locality in memory hierarchy. Some graph benchmarks operating on rich vertex properties, e.g., page rank and triangle count. They incorporate heavy arithmetic computations on vertex properties and intensive memory accesses leading to hybrid workload behaviors. Most vertices are active in all stages of their executions. We selected, compared and studied six common graph benchmarks in this paper. We introduce their details as follows:
\vspace{-0.05in}
\begin{itemize}
\item {\bf Breadth First Search} traverses a graph by starting from a root vertex and exploring its neighbors first.
\item {\bf K-Core} finds a maximal connected sub-graph where all vertices have a $\geq K$ degree. It follows Matula \& Beck's algorithm~\cite{Nai:SC2015}.
\item {\bf Single Source Shortest Path} calculates the minimum cost paths from a root vertex to each vertex in a given graph by Dijkstra's algorithm~\cite{Nai:SC2015}.
\item {\bf Graph Coloring} partitions a graph into independent sets. One set contains vertices sharing the same color. It adopts Luby-Jones' algorithm~\cite{Nai:SC2015}.
\item {\bf Page Rank} uses the probability distribution to compute page rankings. The rank of each vertex is computed by $PR(u)=\sum_{N_u}\frac{PR(v)}{E(v)}$, where the {\it PR} of a vertex ($u$) is decided by the {\it PR} of each vertex ($v$) in its neighbor set ($N_u$) divided by its outgoing edge number $E(v)$.
\item {\bf Triangle Count} measures the number of triangles that are formed in a graph when three vertices are connected to each other (Schank's algorithm~\cite{Nai:SC2015}).
\end{itemize}

\subsection{Graph Topology}

The performance of graph applications is heavily influenced by the topology of graph datasets. We studied how topologies impact the benchmark performance via following metrics. The {\it eccentricity} $\epsilon(v)$ of a vertex $v$ in a given connected graph $G$ is the maximum graph distance between $v$ and any other vertex $u$ of $G$. The {\it diameter} $d$ of a graph is the maximum eccentricity of any vertex in the graph ($d=\max_{v\in V}\epsilon(v)$).  The {\it Fiedler eigenvalue} of the graph ($F(G)$) is the second smallest eigenvalue of the Laplacian matrix of $G$. The magnitude of $F(G)$ exhibits how well $G$ is connected. For traversal-based applications such as breadth first search, traversal iteration number is proportional to the eccentricity and Fiedler eigenvalue. The {\it vertex degree} indicates the number of edges connected to a vertex. The average vertex degree and the vertex degree distribution of a graph reflect the amount of parallelism. 

\subsection{Intel Xeon Phi Architecture}

\subsubsection{Overall Architecture} 
Xeon Phi MIC processors are fully compatible with the x86 instruction set and hence support standard parallel shared memory programming tools, models and libraries such as OpenMP. The KNL~\cite{Sodani:HOT2015} is the 2nd generation architecture (14nm) and has been adopted by several data centers such as the national energy research scientific computing center~\cite{Barnes:PMBS2016}.

\begin{figure}[htbp]
\vspace{-0.1in}
\centering
\includegraphics[width=3.3in]{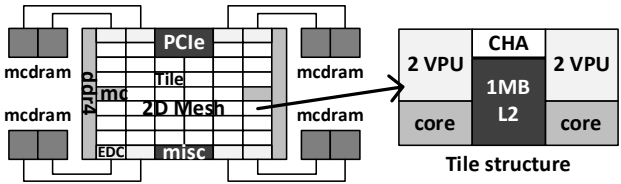}
\vspace{-0.1in}
\caption{KNL (MC: DDR4 controller; EDC: MCDRAM controller).}
\label{fig:f_phi_knl}
\vspace{-0.1in}
\end{figure}

The detailed KNL architecture is shown in Figure~\ref{fig:f_phi_knl}. It is built by up to 72 Atom (Silvermont) cores, each of which is based on a low operating frequency 2-wide issued out-of-order (OoO) micro-architecture supporting four concurrent threads, a.k.a, simultaneous multithreading (SMT). Additionally, every core has two vector processing units (VPUs) that support SIMD instructions such as {\it AVX2} and {\it AVX512} (a new 512-bit advanced vector extension of SIMD instructions for the x86 instruction set). A VPU can execute up to 16 single precision operations or 8 double precision floating point operations in each cycle. Two cores form a tile sharing a 1MB 16-way L2 cache and a caching home agent (CHA), which is a distributed tag directory for cache coherence. All tiles are connected by a 2D Mesh network-on-chip (NoC). 

\begin{figure}[htbp]
\vspace{-0.1in}
\centering
\includegraphics[width=3.3in]{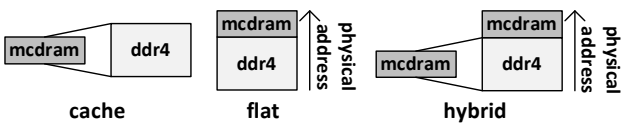}
\vspace{-0.1in}
\caption{Configurations on MCDRAM.}
\label{fig:f_phi_mcdram}
\end{figure}

\subsubsection{MCDRAM} 
The KNL main memory system supports up to 384GB of DDR4 DRAM and 8$\sim$16GB of 3D stacked MCDRAM. The MCDRAM significantly boosts the memory bandwidth, but the access latency of MCDRAM is actually longer than that of DDR4 DRAM~\cite{McCalpin:MCDRAM2016}. As Figure~\ref{fig:f_phi_mcdram} shows, MCDRAM can be configured as a parallel memory component to DDR4 DRAM in main memory ({\it flat} mode), a hardware-managed L3 cache ({\it cache} mode) or both ({\it hybrid} mode). The {\it flat} mode offers high bandwidth by MCDRAM and low latency by DDR4 DRAM. However, programmers have to track the location of each data element and manage software complexity. On the contrary, the {\it cache} mode manages MCDRAM as a software-transparent direct-mapped cache. The {\it hybrid} mode combines the other two modes.   

\begin{figure}[htbp]
\vspace{-0.1in}
\centering
\includegraphics[width=3.3in]{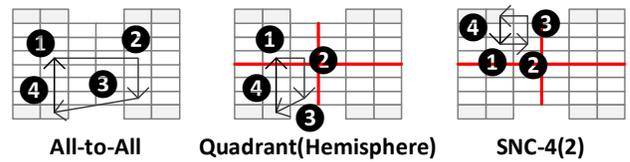}
\vspace{-0.15in}
\caption{Configurations on Cache Clustring.}
\label{fig:f_phi_clustring}
\vspace{-0.15in}
\end{figure} 

\subsubsection{Cache Clustering Mode}
\label{s:intro_ccm}

Since all KNL tiles are connected by a Mesh NoC where each vertical/horizontal link is a bidirectional ring, all L2 caches are maintained coherent by the MESIF protocol. To enforce cache coherency, KNL has a distributed cache tag directory organized as a set of per-tile tag directories (shown as CHA in Figure~\ref{fig:f_phi_knl}) that record the state and location of all memory lines. KNL identifies which tag directory records a memory address by a hash function. As Figure~\ref{fig:f_phi_clustring} shows, the KNL cache can be operated in five modes including {\it All-to-All}, {\it Hemisphere}, {\it Quadrant}, {\it SNC-2} and {\it SNC-4}. When \ding{182} a core confronts a L2 miss, \ding{183} it sends a request to look up a tag directory. \ding{184} The directory finds a miss and transfers this request to a memory controller. \ding{185} At last, the memory controller fetches data from main memory and returns it to the core enduring the L2 miss. In the {\it All-to-All} mode, memory addresses are uniformly hashed across all tag directories. During a L2 miss, the core may send a request to a tag directory physically located in the farthest quadrant from the core. After the directory finds a miss, it may also send this request to a memory controller located in a third quadrant. Therefore, a L2 miss may have to traverse the entire Mesh to read one line from main memory. The {\it Quadrant} ({\it Hemisphere} ) mode divides a KNL chip into four (two) parts. During a L2 miss, the core still needs to send a request to every on-chip tag directory, but the data associated to the target tag directory must be in the same part that the tag directory is located. Memory accesses from a tag directory are managed by its local memory or cache controller. The {\it Hemisphere} and {\it Quadrant} modes are managed by hardware and transparent to Operating System (OS). In contrast, the sub-NUMA-clustering ({\it SNC-2/4}) also separates the chip into two or four parts and exposes each  as a separate NUMA domain to the OS. During a L2 miss, the core only needs to send a request to its local tag directory that also transfers this request to the local memory controller if there is a directory miss. Therefore, the {\it SNC-4} mode has the lowest local memory access latency, but longer memory latency than that of the {\it Quadrant} mode when a request crosses NUMA boundaries. This is because requests traveling to another NUMA region have to be managed by the OS or applications themselves.

\vspace{-0.1in}
\section{Related Work}
\label{s:related_work}

A recent graph characterization work on CPUs~\cite{Beamer:IISWC2015} identifies that eight fat OoO Ivy Bridge cores fail to fully utilize off-chip memory bandwidth when processing graphs, since the small instruction window cannot produce enough outstanding memory requests. In contrast, the KNL MIC processor can saturate main memory bandwidth by $>64$ small OoO cores, each of which supports SMT and has two VPUs. More research efforts~\cite{Burtscher:IISWC2012} analyze GPU micro-architectural details to identify bottlenecks of graph benchmarks. In this paper, we pinpoint inefficiencies on the KNL MIC architecture when running multi-threaded graph applications. Previous physical-machine-based works~\cite{Jiang:ICS2016,Chen:IPDPS2015} find that the first generation Xeon Phi, Knight Corner, has poor performance for graph applications due to its feeble cores. Although the latest simulator-based work~\cite{Ahamd:IISWC2015} characterizes the performance of multi-threaded graph applications on a MIC processor composed of 256 single-issue in-order cores, it fails to consider the architectural enhancements offered by KNLs, e.g., OoO cores, SMT, cache clustering modes, VPUs and MCDRAM. To our best knowledge, this work is the first to characterize and analyze the performance of multi-threaded graph applications on the KNL MIC architecture.

\vspace{-0.05in}
\section{Experimental Methodology}
\label{s:exper_method}
\vspace{-0.05in}

{\bf Graph benchmarks}. In this paper, we adopted and studied six benchmarks including breadth first search (BF), single source shortest path (SS), graph coloring (GC), k-core (KC), triangle count (TC) and page rank (PR) from graphBIG~\cite{Nai:SC2015}. Each benchmark contains a CPU OpenMP version and a GPU CUDA version. Due to the absence of PR GPU code in graphBIG, we created a vertex-centric GPU PR implementation based on a previous GPU PR benchmark~\cite{Burtscher:IISWC2012} and the System-G framework. We compiled codes on CPU by icc (17.0.0) with -O3, -MIC-AVX-512 and Intel OpenMP library. Through the command {\it numactl}, we chose to run CPU benchmarks in DDR4 DRAM under the flat mode. On GPU, we compiled programs by nvcc (V8.0.61) with CUDA-8.0. Since the size of large graph datasets exceeds the maximum physical GPU memory capacity, we used the {\it cudaMallocManaged} function to allocate data structures into a CUDA-managed memory space where both CPUs and GPUs share a single coherent address space. In this way, CUDA libraries transparently manage GPU memory access, data locality and data migrations between CPU and GPU. The OS we used is Ubuntu-16.04.

\begin{table}[htbp]
\centering
\setlength{\tabcolsep}{3pt}
\caption{Benchmark Datasets.}
\vspace{-0.05in}
\begin{tabular}{|l||l|l|l|l|l|}
\hline
\rule{0pt}{9pt}Dataset      & Abbr.   & Vertices  & Edges    &$\overline{Degree}$   & $\overline{Iter_{BF}}$\\ \hline\hline
roadNet\_CA  &\multirow{2}{*}{road}   & 1.97M 	  & 5.53M    & 2.81                 & 665               \\ \cline{1-1}\cline{3-6}
roadNet\_TX  &                        & 1.38M	    & 3.84M    & 2.27                 & 739               \\ \hline
soc-Slashdot0811  &\multirow{2}{*}{social }  & 0.08M	    & 0.9M    & 11.7            & 18                \\ \cline{1-1}\cline{3-6} 
ego-Gplus         &                          & 0.11M 	  & 13.6M    & 127.1                & 6                \\ \hline
delaunay\_n18     &\multirow{2}{*}{delaunay} & 0.26M 	& 1.57M    & 6.0                  & 223               \\ \cline{1-1}\cline{3-6}
delaunay\_n19     &                          & 0.52M	& 3.15M    & 6.0                  & 296               \\ \hline
kron\_g500-logn17 &\multirow{2}{*}{kron}     & 0.11M 	& 10.23M   & 78.0                 & 3                 \\ \cline{1-1}\cline{3-6}
kron\_g500-logn18 &                          & 0.21M	& 21.17M   & 80.7                 & 3                 \\ \hline
twitter7           &\multirow{2}{*}{large}    & 17M  	& 0.48B      & 4.6                  & 23               \\ \cline{1-1}\cline{3-6}
com-friendster  	&                          & 65.6M	& 1.81B    & 27.6                 & 15               \\ \hline
\end{tabular}
\label{t:t_benchmark_dataset}
\vspace{-0.25in}
\end{table}

{\bf Graph datasets}. The graph inputs for our simulated benchmarks are summarized in Table~\ref{t:t_benchmark_dataset}, where $\overline{Degree}$ denotes the average vertex degree of the graph and $\overline{Iter_{BF}}$ describes the average BF iteration number computed by searching from 10K random root vertices. {\it roadNet\_XX} (road) are real-world road networks where most vertices have an outgoing degree below 4. {\it soc-Slashdot0811} and {\it ego-Gplus} (social) are two social networks typically having a scale-free vertex degree distribution and a small diameter. {\it delaunay} (delaunay) datasets are Delaunay triangulations of random points in the plane that also have extremely small outgoing degrees. Like social networks, {\it kronecker} (kron) datasets have large average vertex outgoing degree and can be traversed by only several BF iterations. We also included two large graph datasets (large), {\it twitter7} and {\it com-friendster}, each of which contains millions of vertices and billions of edges. The size of graph datasets is mainly decided by the number of edges, because the average degree of graphs we chose is $\geq2$. We selected graph datasets from the Stanford SNAP~\cite{snapnets}. We also used the synthetic graph generator, PaRMAT~\cite{Khorasani:PACT2015}, in dataset sensitivity studies.

\begin{table}[htbp]
\vspace{-0.1in}
\centering
\caption{Machine Configurations.}
\vspace{-0.05in}
\setlength{\tabcolsep}{3pt}
\begin{tabular}{|l||l|}
\hline
Hardware            & Description \\ \hline\hline
Intel Xeon          & launched in Q2'16, 8-core@2.5GHz, 2T/core,\\
E5-4655v4           & 3.75MB L3/core, 512GB 68GB/s DDR4 DRAM \\ 
                    & thermal design power (TDP) 135W            \\\hline
Nvidia              & launched in Q2'16, 1920-cudaCore@1.5GHz,\\
GTX1070             & peak SP perf: 5783 GFLOPS, 8GB 256GB/s \\ 
                    & GDDR5, PCIe 3.0 $\times$16 to CPU, TDP 150W \\\hline
Nvidia Tesla        & launched in Q3'16, 3840-cudaCore@1.3GHz,\\
P40                 & peak SP perf: 10007 GFLOPS, 24GB 346GB/s \\
                    & GDDR5, PCIe 3.0 $\times$16 to CPU, TDP 250W \\\hline
                    & launched in Q2'16, 64-core@1.3GHz, 4T/core, \\
Intel Xeon Phi      & 0.5MB L2/core, peak SP perf: 5324 GFLOPS, \\ 
7210 (KNL)          & 8 channels 16GB 400GB/s 3D MCDRAM,\\
                    & 512GB 102GB/s DDR4 DRAM, TDP 230W \\\hline
Disk                & 4TB SSD NAND-Flash     \\\hline
\end{tabular}
\label{t:t_hardware_comparison}
\vspace{-0.1in}
\end{table}

\begin{figure*}
\centering
\includegraphics[width=\textwidth]{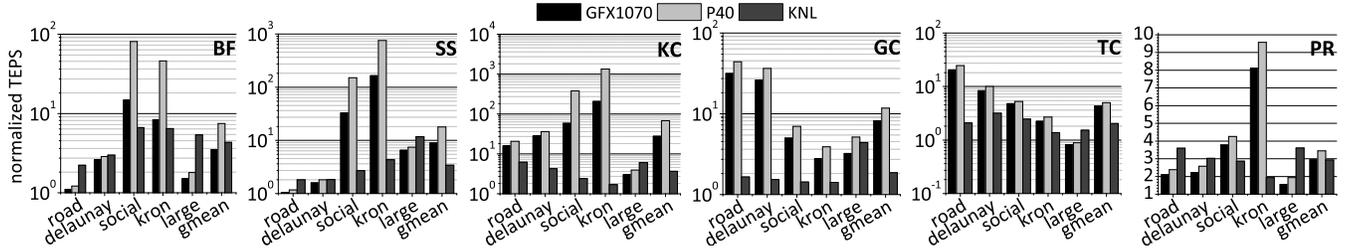}
\vspace{-0.2in}
\caption{Performance comparison between Xeon E5-4655v4, GeForce GTX1070, Tesla P40 and KNL (normalized to Xeon E5-4655v4).}
\label{fig:perf_comp_arch}
\vspace{-0.2in}
\end{figure*}

{\bf Hardware platforms}. We chose and compared three server-level hardware platforms including a Xeon E5 CPU, a Tesla P40 GPU and a Xeon Phi 7210 (KNL) MIC processor. Since, compared to KNL, Tesla P40 achieves almost double peak single point float (SPF) throughput, we also included an Nvidia GTX1070 GPU having similar peak SPF throughput in our experiments. The machine configurations are shown in Table~\ref{t:t_hardware_comparison}. Both Xeon E5 and KNL have a 512GB DDR4 main memory system, while P40 and GTX1070 are also hosted in the Xeon E5 machine with 512GB DDR4 DRAM. The thermal design power (TDP) values of Xeon E5, GFX1040, P40 and KNL are 135W, 150W, 250W and 230W. TDP represents the maximum amount of heat generated by a hardware component that the cooling system can dissipate under any workload.

{\bf Profiling tools}. We adopted Intel VTune Analyzer to collect architectural statistics and likwid-powermeter~\cite{Treibig:PSTI2010} to profile power consumption on CPU and KNL. The likwid-powermeter is a tool for accessing RAPL registers on Intel processors, so it is able to measure the power consumption of both CPU package and DRAM memories. The register reading interval we set in all experiments is $1ms$. For GPUs, we used Nvidia visual profiler ({\it nvprof}) to collect both performance and power characterization results. The performance results we reported in Section~\ref{s:eva} do not include the overhead of profiling tools on CPU, GPU and KNL.

{\bf Metrics}. We measured the performance of graph applications as {\it Traversed Edges Per Second} (TEPS)~\cite{Murphy:CUG2010}. The TEPS for non-traversing-based benchmarks, e.g., PR, is computed by dividing the number of processed edges by the mean time per iteration, since each vertex processes all its edges in one iteration~\cite{Murphy:CUG2010}. We aim to understand the performance and power consumption of multi-threaded graph application kernels, and thus all results ignore disk-to-memory and host-to-device data transfer time during application initializations. But we measured page faults on CPU and data transfers between CPU and GPU inside application kernels.

\section{Evaluation}
\label{s:eva}

We cannot answer which type of hardware is the best for shared memory parallel graph processing, since the platforms we selected have completely different power budgets and hardware configurations such as operating frequencies, core numbers, cache sizes and micro-architecture details. By allocating more power budgets and hardware resources, theoretically speaking, any type of platform can possibly outperform the others. Our purpose of the comparison between KNL and other types of hardware is to demonstrate the KNL MIC processor achieves encouraging performance and power efficiency, and thus it can become one of the most promising alternatives to traditional CPUs and GPUs when processing graphs.

To investigate the KNL performance of parallel graph processing, we empirically analyze the performance comparison between four hardware platforms. And then, we characterize performance details of six graph benchmarks on KNL. We explain the impact of KNL architectural innovations such as many OoO cores, SMT, VPU and MCDRAM on multi-threaded graph applications.

\vspace{-0.1in}
\subsection{Performance Comparison}

The performance comparison between four hardware platforms is shown in Figure~\ref{fig:perf_comp_arch}, where all results are normalized to the performance (TEPS) of Xeon E5 CPU. We configured KNL cache clustering in the Quadrant mode and MCDRAM in the cache mode. The KNL performance reported in this section is achieved by its optimal thread configuration including thread number and thread affinity. We also performed an exhaustive search on all possible configurations on CPU and GPU to attain and report the best performance.

\begin{figure*}
\centering
\includegraphics[width=\textwidth]{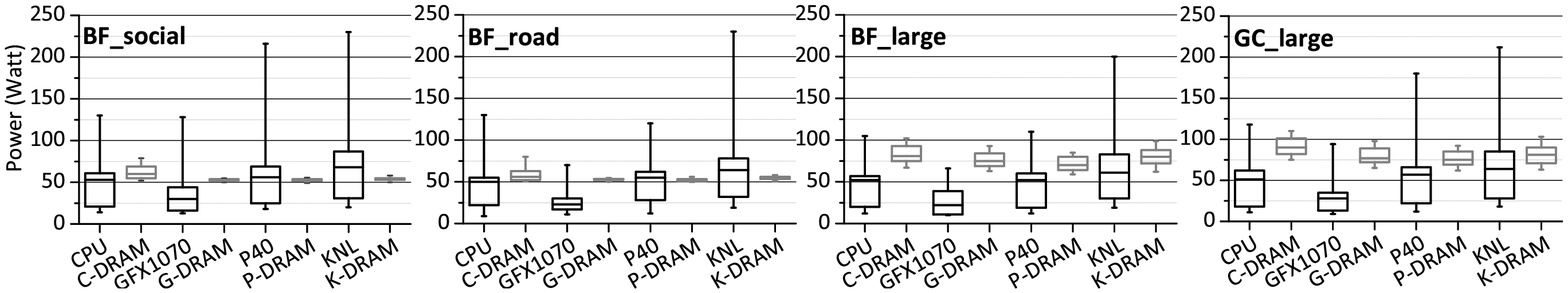}
\vspace{-0.2in}
\caption{Power comparison between Xeon E5-4655v4, GeForce GTX1070, Tesla P40 and KNL.}
\label{fig:knl_power_all}
\vspace{-0.1in}
\end{figure*}

\begin{figure*}
\centering
\includegraphics[width=\textwidth]{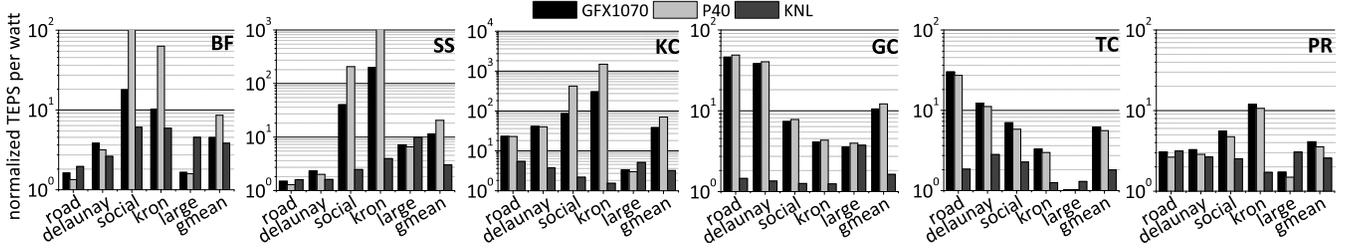}
\vspace{-0.2in}
\caption{Performance per watt comparison between Xeon E5-4655v4, GeForce GTX1070, Tesla P40 and KNL (normalized to Xeon E5-4655v4).}
\label{fig:knl_power_perf}
\vspace{-0.2in}
\end{figure*}



The primary weakness of GPU is the load imbalance problem introduced by its sensitivity to graph topologies in traversal-based graph applications. For BF, GPUs achieve better performance than KNL on graphs with small diameter and large average vertex degree, e.g., social and kron. Huge traversal parallelism exists in these graphs, and hence, the more cores one platform has, the better performance it can achieve. In this case, KNL is slower than GTX1070, although they have similar peak SFP computing throughput. This is because graph traversal operations can barely take advantage of VPUs on KNL. However, compared to KNL, GPUs processes less edges when traversing graphs with large diameter and small average vertex degree, e.g., road and delaunay. There is little traversal parallelism in these graphs, so the OoO cores of KNL prevail due to their powerful sequential execution capability. When processing large graphs, GPUs suffer from frequent data migrations between CPU and GPU memories, due to their limited GDDR5 memory capacity. Therefore, KNL shines on these large graph datasets. P40 outperforms GTX1070 on all graph inputs, because it has more CUDA cores and a larger capacity GDDR5 memory system. As another traversal-based graph benchmark, SS shares the same performance trend as that of BF. 

KC and GC are two applications operating on graph structures. Their computations involve a huge volume of basic arithmetic operations, e.g., integer in/decrementing, on vertices or edges. A large number of GPU cores support these massive arithmetic operations better than a small number of KNL cores. GPUs perform fine-grained scheduling on warps, while each KNL core can switch between only four threads, each of which cannot be fully vectorized. Therefore, the GPU performance is generally 10$\times$ better than that of KNL. For GC, the improvement brought by thousands of CUDA cores on P40 even mitigates the data transfer penalty due to its limited GDDR5 memory capacity, i.e., P40 improves the processing performance over KNL by 16.7\% on large graph datasets. 

As graph benchmarks computing on vertex properties, TC and PR exhibit hybrid workload behaviors, since they require not only graph traversals but also relatively heavy arithmetic operations on vertex properties, e.g., multiply-accumulate operations. In TC, arithmetic operations dominate the performance of graph processing, so compared to KNL, GPUs boost the application performance by 112\%$\sim$145\% averagely. In PR, graph traversals dominate on graphs with large diameter and small average vertex degree, e.g., road and delaunay, so KNL obtains better edge processing speed on these graphs. For almost all combinations of graph benchmarks and datesets, Xeon E5 achieves the worst performance, i.e., only 1\%$\sim$33.9\% of CPU performance or 8\%$\sim$50\% of KNL performance. Only for TC, GPUs are slightly slower than Xeon E5 on large graph datasets, due to large penalties of data transfers on PCIe links between CPU and GPU memories.

\subsection{Power and Performance Per Watt Comparison}

The details of power characterization is shown in Figure~\ref{fig:knl_power_all}, where we list power consumptions of four types of hardware and their 512GB DRAM main memories represented by {\it X-DRAM} (X can be Xeon E5 \{C\}PU, \{G\}FX1070, \{P\}40 and \{K\}NL.). The KNL power includes the power of MCDRAM, while the GPU power contains the power of its GDDR5 memory. 

\begin{figure*}[thbp] 
\centering
\includegraphics[width=\textwidth]{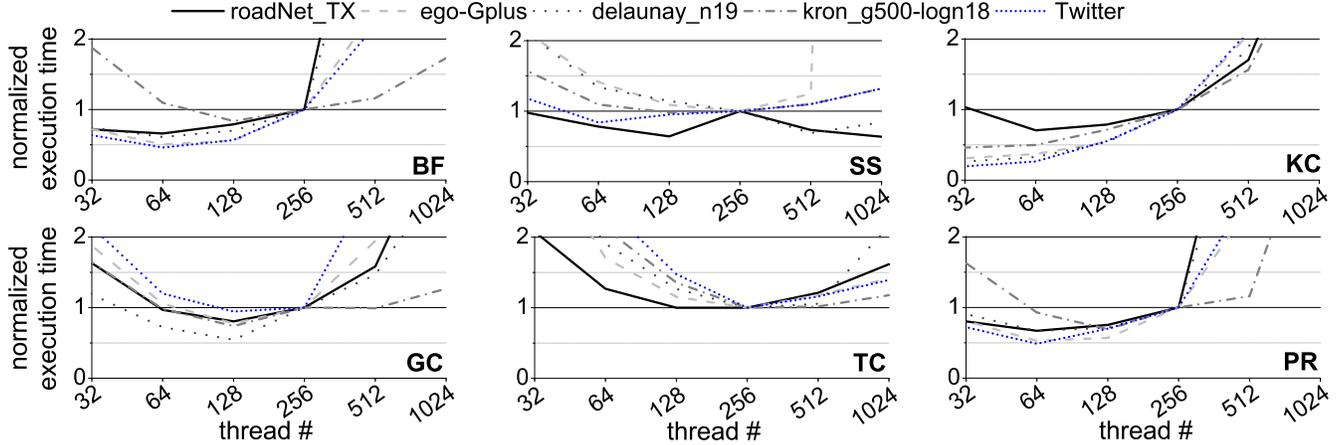}
\vspace{-0.3in}
\caption{The KNL performance with varying thread numbers (normalized to the performance achieved by 256 threads).}
\label{fig:KNL_thread_number}
\vspace{-0.1in}
\end{figure*}

\begin{figure*}[hbtp] 
\vspace{-0.05in}
\centering
\includegraphics[width=\textwidth]{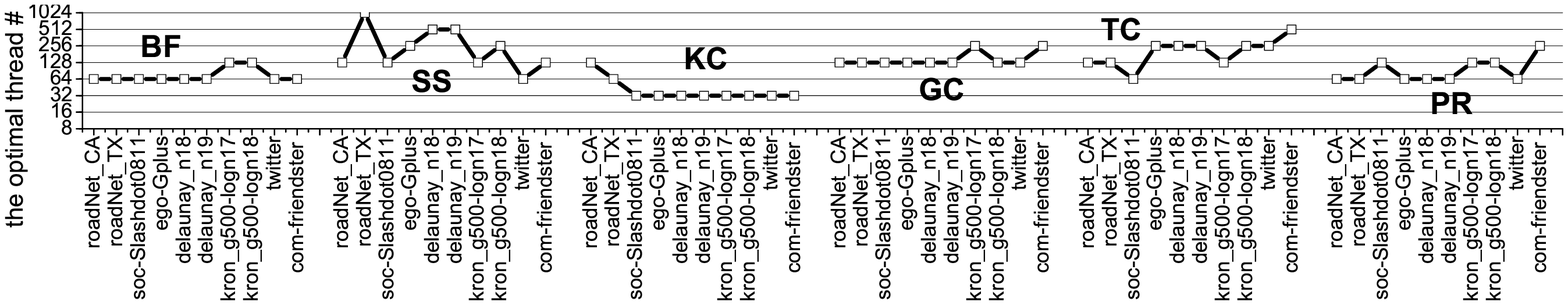}
\vspace{-0.25in}
\caption{The optimal thread number for all benchmarks with all datasets}
\label{fig:perf_opt_thread}
\vspace{-0.25in}
\end{figure*}

\vspace{-0.1in}
For social graph datasets, all platforms are able to fully occupy almost 100\% hardware resources and approach their their peak power in a short time, because large traverse parallelism exists in these datasets. Only CPU DRAM is heavily accessed during BFs, while other platforms barely access their 512GB main memories. This is because both GPUs has their own GPU memories and KNL has a 16GB MCDRAM-based cache. As a result, only CPU DRAM achieves $>60W$ average power consumption. For road graph datasets, the largest power spent by P40 (GFX1070) during searches is only around 48\% (46\%) of its TDP, due to the lack of traverse parallelism in these datasets. The average power consumption of two GPUs when processing road datasets decreases by 16\%$\sim$24\% over that dissipated during traversing on social datasets. In contrast, compared to social graphs, CPU and KNL slightly decrease their average power consumption by only 6\%$\sim$9\%, and the power consumption of CPU DRAM decreases by 10\% averagely when processing road graphs. Compared to GPUs, CPU and KNL are insensitive to topologies of graph datasets due to their limited number of cores. Therefore, they have more consistent power results throughout various graph inputs. For large graph datasets, GPUs barely approach their TDPs due to frequent data transfers through PCIe links. CPU and KNL also suffers from frequent page faults, and thus their largest power values are only around 70\%$\sim$80\% of their TDPs. However, compared to other datasets, DRAMs in all platforms significantly boost the power consumption by 39\%$\sim$62\% when dealing with large graph inputs. SS shares a similar power consumption trend with BF. The power consumption of TC, KC, GC and PR with small datasets is similar to BF with social datasets, since all platforms enjoy similarly large graph processing parallelism and small datasets do not need to frequently visit DRAMs. But when running TC, KC, GC and PR with large datasets, DRAMs spend 4\%$\sim$11\% more power than that when processing large datasets in BF. This is because graph applications operating on both structures of large graphs and a huge volume of vertex properties send more intensive memory accesses to DRAMs than traversal-based benchmarks.

The performance per watt comparison among all hardware platforms is exhibited in Figure~\ref{fig:knl_power_perf}, where all results are normalized to the performance per watt result of Xeon E5 CPU. Compared to CPU, P40 (GFX1070) achieves 2.5$\sim$69$\times$ (3$\sim$37$\times$) better performance per watt averagely. On average, KNL obtains only 60\%$\sim$285\% better performance per watt over CPU. But for large graph datasets, compared to GPUs, KNL improves the performance per watt by 74\%$\sim$186\% when running BF and PR, and has similar results on performance per watt when running other benchmarks. Frequent data transfers between GPU and CPU memories caused by the limited capacity of GDDR5 memory waste a large amount of power and slow down applications on GPUs when processing large graphs. When running GC, TC and PR, although P40 is faster than GFX1070, the power consumption of P40 is also larger. Therefore, the performance per watt of P40 is not significantly better than that of GFX1070 for these benchmarks.

\vspace{-0.1in}
\subsection{Threading}

\subsubsection{Thread Scaling}

We show the graph application performance with varying OpenMP thread numbers on KNL in Figure~\ref{fig:KNL_thread_number}, where all performance results are normalized to the performance achieved by 256 threads. 256 (4 threads $\times$ 64 cores) is the KNL logic core number. Due to the limited figure space, we selected only the larger dataset in each graph input category. With a small number of threads ($<32$), there are not enough working threads to do the task, and thus no benchmark obtains its best performance. When having more threads, the performance of almost all benchmarks improves more or less. However, when the thread number goes beyond $256$, the execution time significantly rises, since the thread synchronization overhead dominates and degrades the performance of most benchmarks. The peak benchmark performance is typically achieved by $32\sim512$ threads.

SMT and oversubscription allow multiple independent threads of execution on a core to better utilize hardware resources on that core. To implement SMT, some hardware sections of the core (but not the main execution pipeline) are duplicated to store architectural states. When one thread is stalled by long latency memory accesses, SMT stores its state to backup hardware sections and switches the core to execute another thread. SMT transforms one physical KNL core to four logic cores, each of which supports one thread. Some applications with certain datasets, e.g., TC with ego-Gplus and SS with kron\_g500-logn18, fulfill their best performance by SMT (256 threads). In contrast, oversubscription requires the assistance from software such as OS or OpenMP library to switch threads, when the running thread is stalled. For applications suffering from massive concurrent cache misses, like SS with {roadNet\_TX} and {delaunay\_n19}, oversubscription supports more simultaneous threads and outruns SMT. When running these applications, compared to the penalty of long latency memory accesses, the OS context switching overhead is not significant.

\begin{figure}[hbtp] 
\centering
\vspace{-0.1in}
\includegraphics[width=3.4in]{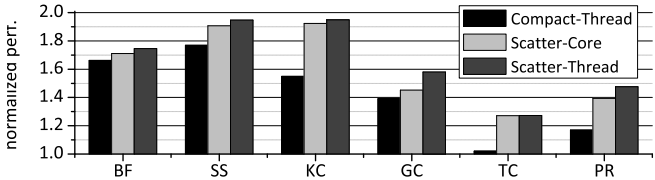}
\vspace{-0.25in}
\caption{Performance comparison between various configurations of thread placement and affinity (normalized to Compact-Core).}
\label{fig:perf_knl_aff}
\vspace{-0.1in}
\end{figure}


\subsubsection{The Optimal Thread Number}
We define the optimal thread number as the thread number achieving the best performance for each benchmark. Figure~\ref{fig:perf_opt_thread} describes the optimal thread number for all applications with all datasets. There is no universal optimal thread number, e.g., the physical core number or the logic core number, that can always have the best performance for all applications with all datasets. Different applications require distinctive optimal thread numbers for their own best performance. Moreover, even for the same application, the optimal thread numbers for various graph datasets are different. As Figure~\ref{fig:KNL_thread_number} shows, na\"ively setting the thread number to 256 decelerates KC with twitter7 by $4.1\times$.

\begin{figure}[hbtp] 
\centering
\vspace{-0.15in}
\includegraphics[width=3.4in]{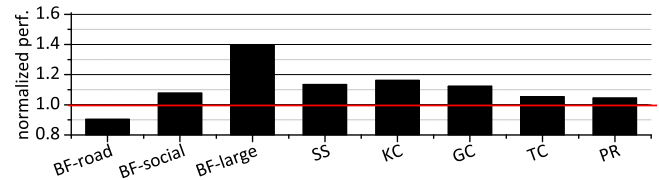}
\vspace{-0.2in}
\caption{Performance comparison between DDR4 and MCDRAM (normalized to DDR4 DRAM main memory without MCDRAM).}
\label{fig:perf_knl_mcdram}
\vspace{-0.1in}
\end{figure}

\subsubsection{Thread Placement and Affinity}

Because graphBIG depends on the OpenMP library, we can configure the thread placement and affinity by {\it KMP\_AFFINITY=X, granularity=Y}. Here, $X$ indicates thread placement and has two options: assigning thread $n+1$ to an available thread context as close as possible to that of thread $n$ (Compact) or distributing threads as evenly as possible across the entire system (Scatter). And $Y$ denotes granularity and includes two choices: allowing all threads bound to a physical core to float between different thread contexts (Core) or causing each thread to be bound to a single thread context (Thread). The performance comparison between all configurations of thread placement and affinity is shown in Figure~\ref{fig:perf_knl_aff}, where each bar represents one {\it X-Y} combination and all bars are normalized to Compact-Core. We see that Compact configurations with Core and Thread have worse performance, since Scatter configurations better utilize all physical cores and distribute memory requests evenly among all memory controllers. In two Scatter configurations, granularity Thread wins slightly better performance, because it scatters consecutive threads sharing similar application behaviors to different physical cores. 

\subsection{MCDRAM}

We configured MCDRAMs as a hardware-managed L3 cache for KNL as default. However, we can also disable MCDRAMs to use only DDR4 DRAM as main memory. The performance improvement of MCDRAM is exhibited in Figure~\ref{fig:perf_knl_mcdram}, where all results are normalized to the performance of KNL with only DDR4 DRAM main memory. Compared to DDR4, MCDRAM can supply $4\times$ bandwidth. The MCDRAM-based cache boosts the performance of most benchmarks by its 16GB capacity and larger bandwidth. Particularly, large graph datasets benefit more from the MCDRAM-based cache, since they enlarge the working set size for most applications. On the contrary, the access latency of MCDRAM is longer than that of DDR4 DRAM~\cite{McCalpin:MCDRAM2016}. Some benchmarks processing graphs with large diameter and small average degree, e.g. BF with road datasets, are more sensitive to the prolonged memory access latency and actually decelerated by the MCDRAM-based cache, because they have a limited number of concurrently pending memory accesses and small memory footprints.

\begin{figure}
\centering
\includegraphics[width=3.4in]{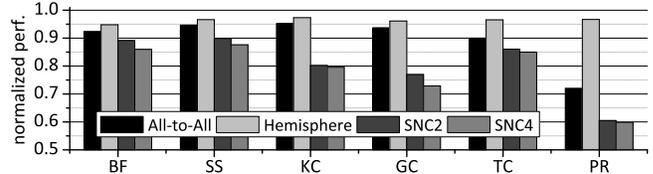}
\vspace{-0.2in}
\caption{Performance comparison between different cache clustering modes (normalized to {\it Quadrant}).}
\label{fig:perf_imp_noc}
\vspace{-0.2in}
\end{figure}

\subsection{Vectorization}

We compiled graph programs with icc -O3 with various vectorization choices: no vectorization, {\it AVX2} and {\it AVX512}. {\it AVX2} improves the graph processing performance by 47\%$\sim$324\% over {\it NOVEC}. However, compared to {\it AVX2}, {\it AVX512} does not significantly further boost the performance of graph applications. This is because the implementation of System G does not explicitly represent vertices or edges by floating point or integer arrays. Instead, vertices and edges are encapsulated into lists or maps in graph frameworks, and thus it is difficult to vectorize these data structure on KNL. Moreover, most graph datasets we used are sparse, so they can barely be improved by wide {\it AVX512} SIMD instructions.

\begin{figure*}[hbtp] 
\centering
\subfigure[Vertex (normalized to 100K).]{
\includegraphics[width=2.3in]{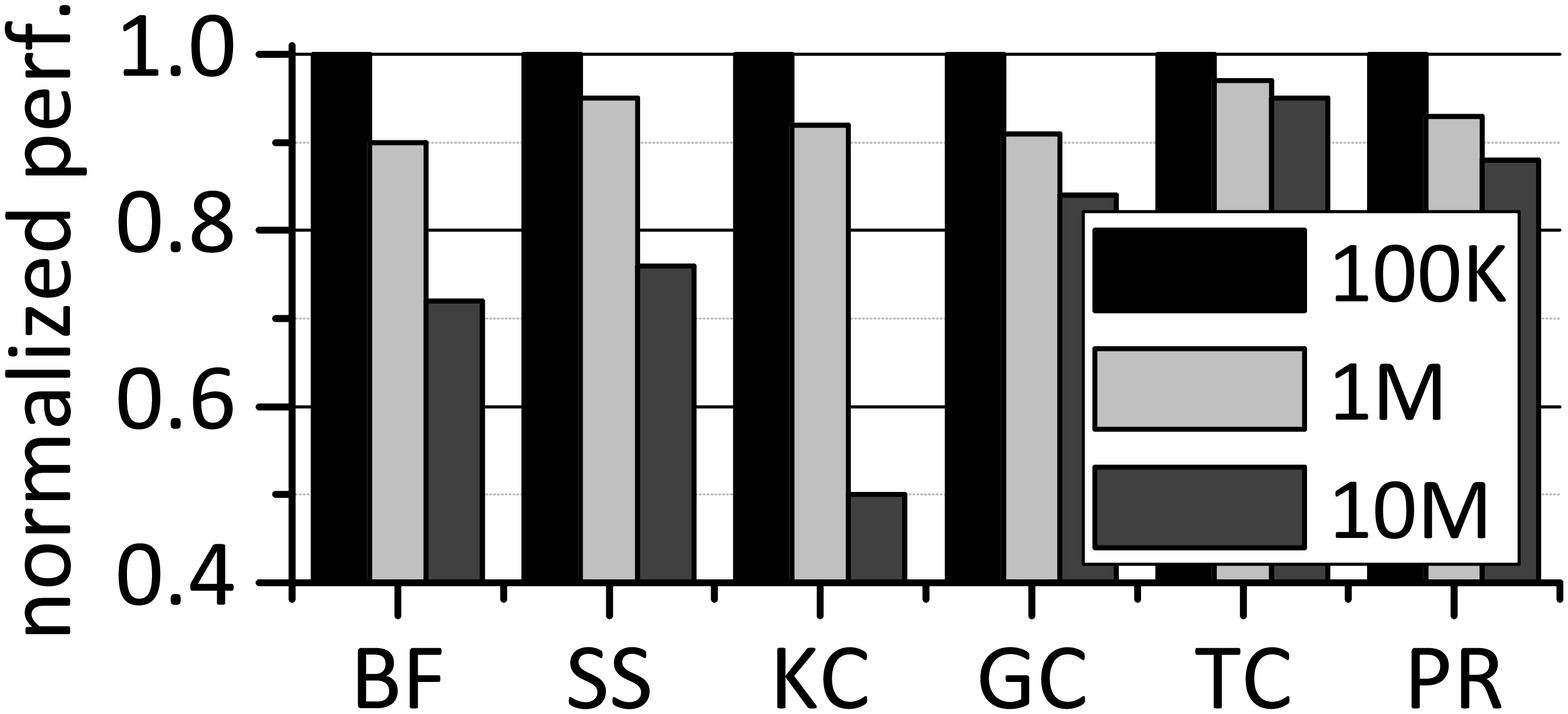}
\label{fig:f_knl_vertex}
}
\hspace{-0.2in}
\subfigure[Degree (normalized to 10).]{
\includegraphics[width=2.3in]{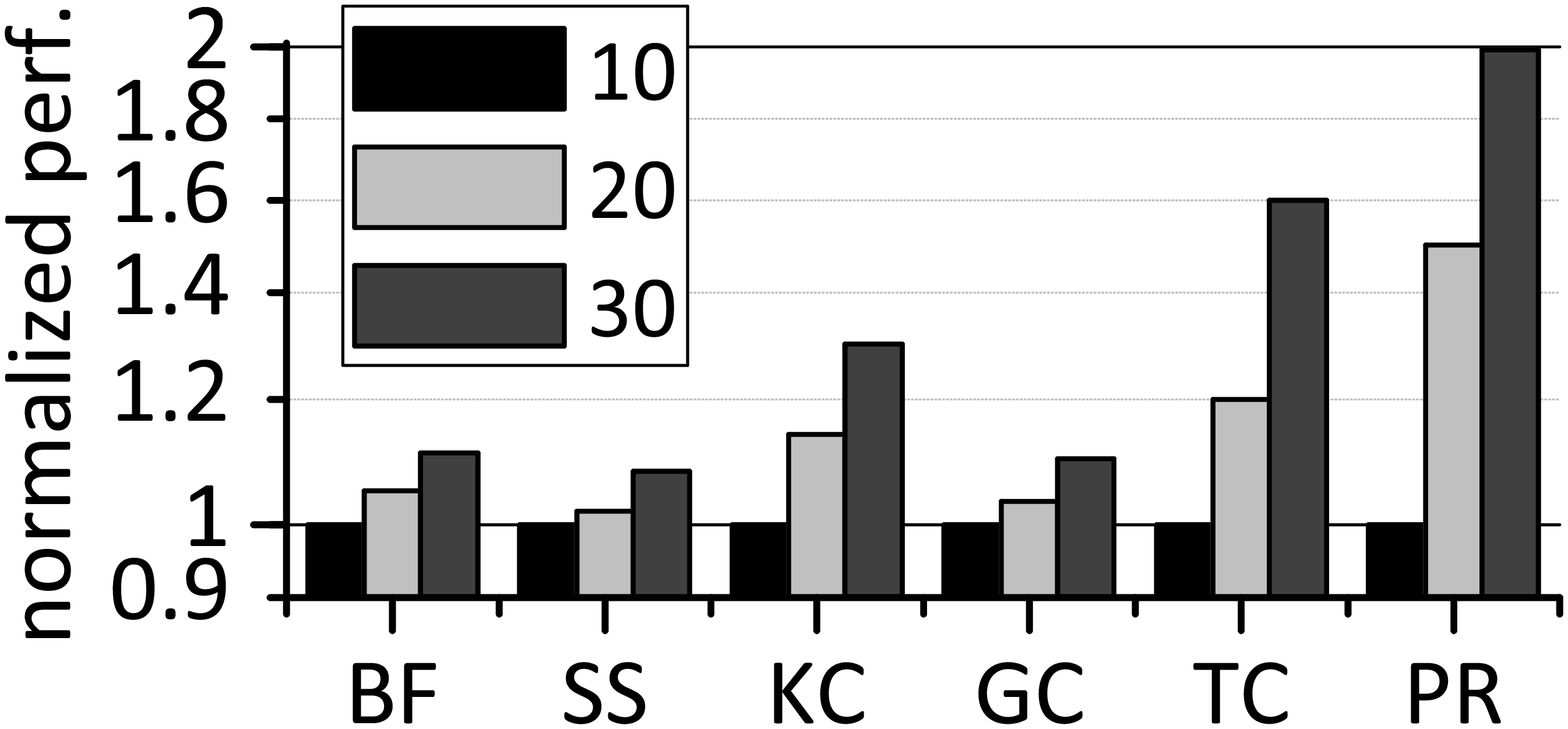}
\label{fig:f_knl_edge}
}
\hspace{-0.2in}
\subfigure[Skewness (normalized to Erd\H{o}s-R\'enyi).]{
\includegraphics[width=2.3in]{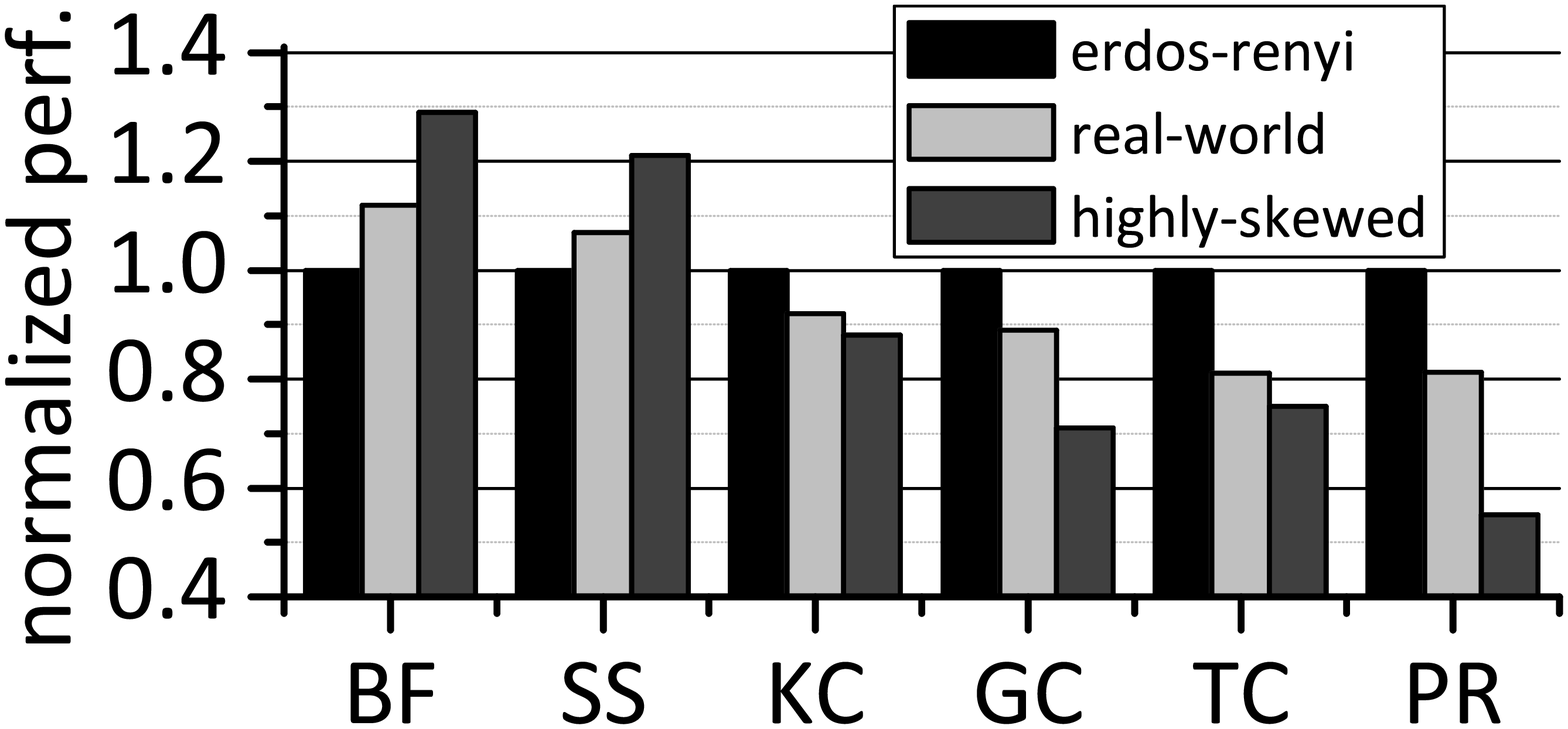}
\label{fig:f_knl_skew}
}
\vspace{-0.1in}
\caption{The KNL performance with various R-MAT datasets.}
\label{fig:perf_data_scaling}
\vspace{-0.2in}
\end{figure*}

\subsection{Cache Clustering Mode}

The performance comparison between various cache clustering modes is shown in Figure~\ref{fig:perf_imp_noc}, where all results are normalized to Quadrant. As explained in Section~\ref{s:intro_ccm}, among all hardware-managed modes including {\it All-to-All}, {\it Hemisphere} and {\it Quadrant}, Quadrant achieves the best performance, since it can keep both L2 accesses and memory accesses served within the local quadrant on KNL. Two software-managed modes ({\it SNC-2} and {\it SNC-4}) have even worse graph processing performance, since benchmarks in graphBIG do not have NUMA-awareness and have to pay huge penalty for frequent communications between different NUMA regions. SNC-4 offers four NUMA regions, therefore, compared to SNC-2, it degrades the graph application performance more by expensive communications between NUMA regions.

\begin{figure}[ht]
\vspace{-0.1in}
\centering
\includegraphics[width=3.3in]{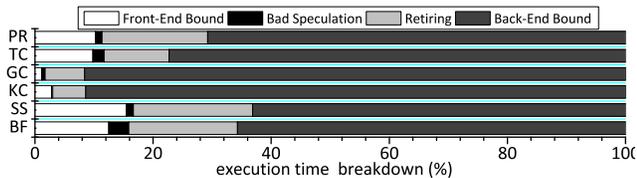}
\vspace{-0.15in}
\caption{KNL execution time breakdown (with twitter7).}
\label{fig:perf_knl_break}
\vspace{-0.2in}
\end{figure}

\subsection{Execution Time Breakdown}

To understand bottlenecks of applications, we show KNL execution time breakdown of all benchmarks with twitter7 graph input in Figure~\ref{fig:perf_knl_break}. {\it Bad Speculation} is the time stall due to branch mis-prediction. {\it Retiring} denotes the time occupied by the execution of useful instructions. {\it Front-End Bound} indicates the time spent by fetching and decoding instructions, while {\it Back-End Bound} means the waiting time due to a lack of required resources for accepting more instructions in the back-end of the pipeline, e.g., data cache misses and main
memory accesses. It is well known that graph applications are extremely memory intensive and have irregular data accesses. The breakdown of execution time on KNL also supports this observation. For KC and GC, $>90\%$ execution time is used to wait for back-end stalls. Although in Figure~\ref{fig:perf_knl_mcdram}, the 16GB MCDRAM-based cache improves the performance of KC and GC by $>10\%$ on average, we anticipate a larger capacity MCDRAM-based cache can further boost their performance, especially when processing large graph datasets.


\subsection{Dataset Sensitivity Analysis}

We generated scale-free graphs (R-MAT~\cite{Chakrabarti:ICDM2004}) with varying numbers of vertices, average vertex degrees and vertex distributions by PaRMAT generator~\cite{Khorasani:PACT2015}. The dataset sensitivity studies on KNL are shown in Figure~\ref{fig:perf_data_scaling}. Among four R-MAT parameters ($a$, $b$, $c$ and $d$)~\cite{Chakrabarti:ICDM2004}, we always enforced $d=0.11$ and $b=c$ for symmetry. To produce different skewnesses, we set the ratio ({\it skewness}) between $a$ and $b$ to 1 (Erd\H{o}s-R\'enyi model), 3 (real-world) and 8 (highly-skewed). In Figure~\ref{fig:f_knl_vertex}, we fixed the average vertex degree to 20 and the {\it skewness} to 3. With an increasing number of vertices, the performance of all benchmarks degrades more or less, mainly because D-TLB and L2 miss rates are increased by larger graph sizes. We fixed the vertex number to 1M and the {\it skewness} to 3 in Figure~\ref{fig:f_knl_edge}. All applications demonstrate better performance with an enlarging average vertex degree, since more edges per vertex lead to increased L2 hit rate. Particularly, the performance of TC and PR increases more obviously, since they operate on neighbor sets in vertex properties and higher vertex degree brings more accesses within vertices. In Figure~\ref{fig:f_knl_skew}, we set the number of vertices and the average vertex degree to 1K and 20 respectively. And we explored the {\it skewness} among 1, 3 and 8. For traversal-based applications, BF and SS, as graph skewness increases and graph eccentricity decreases, the application performance increases. Higher-skewed graphs have smaller diameter resulting in faster traversals. On the contrary, the other graph benchmarks suffer from severer load imbalance, when their datasets are more skewed.



\vspace{-0.1in}
\section{Conclusion and Future Work}
\label{s:con}

In this paper, we present a performance and power characterization study to show the potential of KNL MIC processors on parallel graph processing. To fully utilize KNLs, in future, we need to overcome challenges from both hardware angle and software perspective.

First, from hardware angle, KNL supplies many architectural features that can be configured by knobs. Different graph applications may favor different configurations. For instance, different graph benchmarks require distinctive numbers of threads to achieve the best performance. Furthermore, some applications benefit from high bandwidth MCDRAM, others may be improved by low latency DDR4 DRAM. Therefore, it is vital to have auto-tuning tools to search the optimal configuration of these knobs to achieve the best performance on KNLs. Previous works propose exhaustive iteration-based optimizations~\cite{Kulkarni:TACO2009} and machine-learning-based tuning techniques~\cite{Tiwari:IPDPS2009}. For KNLs, we believe that future auto-tuning schemes have to consider both the MIC architecture and the heterogeneous main memories.

Second, from software perspective, though a state-of-the-art multi-threaded graph framework fully optimized for traditional multi-core CPUs can run on KNLs, we observe that hardware resources such as VPUs are underutilized and advanced software-managed architectural features, e.g., the SNC-4 cache clustering mode, may even hurt the performance of graph applications. In future, instead of optimizing a single benchmark, we need to create a fully vectorized graph framework offering {\it AVX512} friendly primitives to support a wide variety of graph applications on KNLs. Moreover, we should incorporate a OS-based~\cite{Li:SC2007} or library-based~\cite{Li:HPDC2013} NUMA-aware memory management technique into future graph frameworks, so that the graph applications can benefit from the lowest local memory access latency provided by SNC-4 without incurring large communication overhead between NUMA regions. The future graph frameworks on KNLs also need to be rewritten with heterogeneous memory supporting libraries such as MEMKIND to allocate latency sensitive pages to DDR4 DRAMs and bandwidth sensitive pages to MCDRAMs.

\section*{Acknowledgment}
We gratefully acknowledge the support from Intel Parallel Computing Center (IPCC) grant, NSF OCI-114932 and CIFDIBBS-143054. We appreciate the support from IU PHI, FutureSystems team, and ISE Modelling \& Simulation Lab.

\bibliographystyle{short}

\let\oldthebibliography=\thebibliography
\let\endoldthebibliography=\endthebibliography
\renewenvironment{thebibliography}[1]{%
   \begin{oldthebibliography}{#1}%
     \setlength{\itemsep}{-.1ex}%
}%
{%
   \end{oldthebibliography}%
}

\bibliography{graph}

\end{document}